%% file: main.tex
\documentclass[12pt]{article}
\usepackage[utf8]{inputenc}
\DeclareUnicodeCharacter{2212}{-}
\usepackage{graphicx, authblk}
\usepackage{amsfonts,amsmath,amssymb,mathtools}
\usepackage[left=2cm,right=2cm,top=2cm,bottom=2cm,bindingoffset=0cm]{geometry}
\usepackage{cite}
\usepackage{stmaryrd}
\usepackage{tabularx}
\usepackage[dvipsnames]{xcolor}
\usepackage{mathrsfs}
\usepackage{esvect}
\usepackage{fmtcount}
\usepackage{subcaption}
\usepackage[scr=boondox]{mathalpha}
\usepackage{esint}
\usepackage{setspace}
\usepackage{braket}
\usepackage[labelfont=bf, margin=20pt,font={small}]{caption}
\usepackage{subcaption}
\usepackage{youngtab}
\usepackage{epigraph}
\usepackage{upgreek}
\usepackage[english]{babel}
\usepackage[toc,page]{appendix}
\usepackage[page]{appendix}

\newcommand\beq{\begin{equation}}
\newcommand\eeq{\end{equation}}

\usepackage{import}
\usepackage{xifthen}
\usepackage{pdfpages}
\usepackage{transparent}

\graphicspath{{figures/}}

\definecolor{BlueGreen}{RGB}{49,152,255}
\definecolor{Violet}{RGB}{120,80,120}
\definecolor{Blue}{RGB}{0,0,255}
\definecolor{Yellow}{RGB}{0,255,51}
\definecolor{ElectricGreen}{RGB}{0, 255, 0}
\definecolor{MediumPersianBlue}{RGB}{0, 103, 165}
\definecolor{gray}{RGB}{0, 95, 95}

\input{Defs}

\usepackage[unicode, colorlinks, urlcolor=gray, linkcolor=gray, citecolor=Blue]{hyperref}
\numberwithin{equation}{section}



\title{\bf Lessons from the Klein paradox}

\author[1,2]{E.~T.~Akhmedov\thanks{\href{mailto:akhmedov@itep.ru}{akhmedov@itep.ru}}}
\author[1,4]{D.~V.~Diakonov\thanks{\href{mailto:dmitrii.dyakonov@phystech.edu }{dmitrii.dyakonov@phystech.edu}}}
\author[1]{V.~I.~Lapushkin\thanks{\href{mailto:volapushkin@gmail.com}{volapushkin@gmail.com}}}
\author[1,3]{D.~I.~Sadekov\thanks{\href{mailto:sadekov.di@phystech.edu}{sadekov.di@phystech.edu}}}
\affil[1]{\itshape Institutskii per, 9, Moscow Institute of Physics and Technology, 141700, Dolgoprudny, Russia}
\affil[2]{\itshape Academician Kurchatov Square, 1, NRC ''Kurchatov Institute'', 123182, Moscow, Russia}
\affil[3]{\itshape P. N. Lebedev Physical Institute, Moscow 119991, Russia}
\affil[4]{ Institute for Information Transmission Problems }
\date{\today}

\begin{document}
\maketitle

\begin{abstract}
   We re-examine the Klein paradox from a many-particle perspective in quantum field theory. Specifically, we compute the expectation value of the particle current induced by a sufficiently strong step-like electric potential in 1+1 dimensions. First, for a constant (eternal) potential, we calculate the current for different Fock space ground states corresponding to distinct mode bases. While one basis yields a zero current, another produces the standard nonzero value. We then consider a potential that is rapidly switched on, recovering the standard current in the asymptotic future. This result is generalized to potentials that interpolate between different constant values at spatial infinity. Finally, we analyze a potential acting for a finite duration and again reproduce the standard current. A physical interpretation of these results is provided.
\end{abstract}

\newpage

\section{Introduction}\label{sec:Introduction}
The Klein paradox \cite{Klein} (see, e.g., \cite{Calogeracos:1999yp}) is a well-known phenomenon concerning the behavior of fermions in an external electric field. It has been extensively analyzed within both many-particle \cite{Calogeracos:1998rf, Gavrilov:2015yha, Nikishov:1970br, Nikishov1985} and one-particle frameworks \cite{Nikishov:1970br, ternov2024paradoks15483}. The defining feature of this phenomenon from the many-particle perspective is the emergence of a nonzero electric current in the absence of external charges, resulting from a vacuum ``breakdown'' in sufficiently strong electric fields. For example, in a parallel-plate capacitor, this effect occurs when the field strength $E$ and the plate separation $L$ satisfy $eEL > 2mc^2$, where $m$ is the fermion mass (below we use units where $c = 1$ and $\hbar = 1$). To simplify the study of this effect, it is common to work in $d=1+1$ spacetime dimensions and model the electric field as localized, e.g., $E\sim \delta(x)$; this paper adopts the latter simplifications.

From the single-particle perspective, the Klein paradox is that certain solutions of the Dirac equation exhibit a negative transmission coefficient in the ``Klein zone'' of the spectrum. In this zone, the incident and reflected components of an incoming wave have co-directed current and momentum, whereas for the transmitted component, the current and momentum point in opposite directions (see \cite{ternov2024paradoks15483} and Appendix \ref{app:1_particle_solution}). This paradox finds resolution in both many-particle and single-particle interpretations. The many-particle picture describes it as the disappearance of a particle state and the creation of an antiparticle state with opposite charge and current. In the single-particle approach, the paradox can be resolved by invoking the concept of ``backward motion'' \cite{ternov2024paradoks15483}; by an appropriate choice of the Dirac wavefunction, the probability current can be made to point in the natural direction everywhere: the incident and transmitted waves are co-directed and the transmission coefficient is positive, which resolves the paradox.

At the same time, the behavior of quantum fields and the vacuum structure in the presence of strong background fields remain areas of active research, with many open questions. For instance, quantum fields in the early Universe and the gravitational collapse of a star into a black hole, along with subsequent Hawking radiation, are not yet fully understood from the perspective of quantum field theory for general initial states (see, e.g., \cite{Akhmedov:2021rhq} for a short recent review). The goal of the present paper is to address these questions systematically, beginning with the fundamental model problem of the Klein paradox. In this paper, we examine three specific scenarios: a constant electric field (Sec.\ref{sec:Constant electric field}), the process of switching on the electric field (Sec.\ref{sec:Turning on the electric field} and \ref{Lapushkin}), and a finite-duration external field (Sec.\ref{sec:Finite action of the external electric field}).

For a constant field, we demonstrate that the resulting current depends critically on the choice of the basis of modes. Notably, we show that there exists a Fock space ground state—corresponding to a particular choice of modes—for which the current vanishes. For a switched-on field, one can single out a particular basis of modes on physical grounds (that define the Poincar'e-invariant vacuum as the Fock space ground state) before the background field is switched on. We compute the asymptotic future current for such a state. For a finite duration of the background field, we evaluate the current by calculating the total charge of particles produced during the field's action, divided by its duration.

In problems with external background fields, the definitions of ``particles'' and ``antiparticles'' are generally ambiguous, as these concepts inherently refer to asymptotic plane waves with definite momentum. It is more precise to discuss quantum field states and their Hamiltonian evolution. However, for spatially localized external fields, the incoming and transmitted waves are asymptotically plane, allowing them to be associated with particles. Accordingly, in Sec.\ref{sec:Constant electric field}, we introduce creation operators for ``particles'' and ``antiparticles''—related by a CP transformation—and define the corresponding Fock space ground state to represent ``empty'' space. In Sec.\ref{sec:Turning on the electric field} and \ref{Lapushkin}, where the field is switched on, as the initial state prior to the activation of the field, we choose the Poincar'e-invariant vacuum. Finally, in Sec.\ref{sec:Finite action of the external electric field}, where no external fields are present in the asymptotic past and future, we can unambiguously count the number of particles and antiparticles created during the field's action period.

\section{Constant electric field}\label{sec:Constant electric field}

As stated in the Introduction, we first study the problem in a time-independent, i.e., eternally existing, electric field in $2D$. We choose the gauge for the 4-potential as:
\begin{align}
    -eA_{\mu}(x) = V_0\theta(x)\delta_{0\mu},
\end{align}
where: $\mu\in\{0,1\}$, $V_0>2m$ and $e>0$ is the electron charge. Thus, the electric field is localized at the origin:
\begin{align}
    eE(x)= V_0\delta(x).
\end{align}
To quantize the fermion field, one needs the modes solving the Dirac equation:
\beq\label{eq:dirac}
    \Big(i\gamma^{\mu}\partial_{\mu}-(-e)\gamma^{\mu}A_{\mu}-m\Big)\psi(t,x)=0,
\eeq
where:
\beq\label{eq:gamma_choice}
    \gamma^{0} = \begin{pmatrix}
	1 & 0\\
	0 & -1
	\end{pmatrix}, \; \gamma^{1} = \begin{pmatrix}
	0 & i\\
	i & 0
	\end{pmatrix}.
\eeq 
Solutions of (\ref{eq:dirac}) with fixed frequency are given in the form\footnote{To clarify the expressions below, in Appendix A we present the modes for the fermion field in empty 2D Minkowski spacetime in the form that is used here.}:
\begin{align}
    \psi_{\omega}(t,x)=e^{-i\omega t}\psiw(x) = e^{-i\omega t}\begin{pmatrix}
    \psi_L(x)\\
    \psi_R(x)
    \end{pmatrix}.
\end{align}
In terms of the components $\psi_L,\psi_R$, the equation (\ref{eq:dirac}) reads
\begin{equation}\label{eq:dirac_components}
	\begin{cases}
	\big(\omega - V_0\theta(x)-m\big)\psi_L(x) - \partial_x \psi_R(x) = 0,\\
	\big(\omega - V_0\theta(x)+m\big)\psi_R(x) + \partial_x \psi_L(x) = 0.
	\end{cases}
\end{equation}    
\begin{figure}[t]
    \centering
    \includegraphics[width=0.65\linewidth]{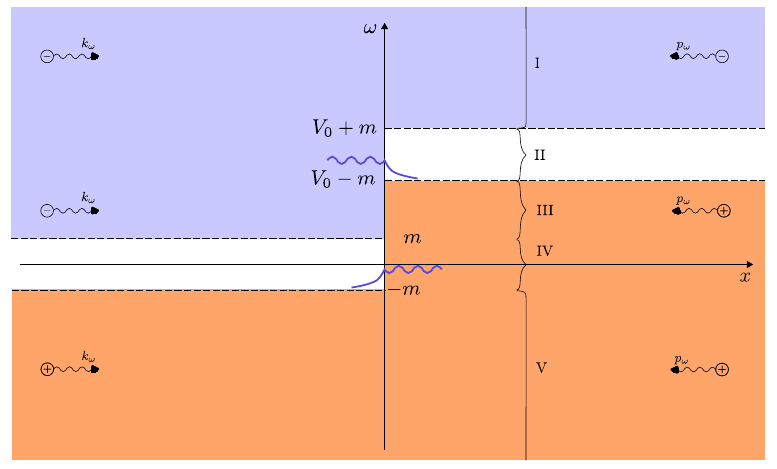}
    \caption{Illustration of the in-modes.}
    \label{Fig:In_modes_plot}
\end{figure}

For each frequency $\omega$, one can write solutions in the form of waves ``incoming'' from infinity that satisfy the continuity condition at the point $x=0$:
\beq\label{eq:in_modes_left}
\begin{aligned}
    &\psiwL(x) = A_{\omega}\Bigg\{ \theta(-x)\frac{2\kappaw\pw\chi(\omega)}{\kw\left(\kappaw+1\right)}
    \begin{pmatrix}
        -i\kw \\
        \omega -m
    \end{pmatrix}e^{-i\kw x} \;+
    \\
    &\quad+\;\theta(x)\Bigg[
    \begin{pmatrix}
        -i\pw\chi(\w)\\
        \omega-V_0 -m
    \end{pmatrix}e^{-i\pw \chi(\w)x} - \frac{\kappaw-1}{\kappaw+1} \begin{pmatrix}
        i\pw\chi(\w)\\
        \omega-V_0 -m
    \end{pmatrix}e^{i\pw\chi(\w) x}\Bigg]\Bigg\}\times\big(1-\theta_{\II}(\omega)\big),
\end{aligned}
\eeq
\beq\label{eq:in_modes_right}
\begin{aligned}
    \psiwR(x) = B_{\omega}\Bigg\{ \theta(-x)\Bigg[
    &\begin{pmatrix}
        i\kw\\
        \omega -m
    \end{pmatrix}e^{i\kw x} + \frac{\kappaw-1}{\kappaw+1} \begin{pmatrix}
        -i\kw\\
        \omega -m
    \end{pmatrix}e^{-i\kw x}\Bigg] \;+
    \\
    &\quad\quad\;\;+\;\theta(x)\frac{2\kw\chi(\omega)}{\pw\left(\kappaw+1\right)}
    \begin{pmatrix}
        i\pw\chi(\omega)\\
        \omega-V_0 -m
    \end{pmatrix}e^{i\pw\chi(\omega)x}\Bigg\}\times\big(1-\theta_{\IV}(\omega)\big),
\end{aligned}
\eeq
where $A_{\omega}$ and $B_{\omega}$ are normalization coefficients to be fixed below, and we introduced the notations
\beq\label{eq:notations}
    \kw = \sqrt{\omega^2-m^2},\;\;\pw = \sqrt{(\omega-V_0)^2-m^2},\;\; \kappaw = \frac{\kw}{\pw}\frac{\omega-V_0-m}{\omega-m}\chi(\w),
\eeq
together with several ``characteristic'' functions of frequency regions marked in Fig.\ref{Fig:In_modes_plot}:
\beq\label{eq:theta_functions}
\begin{aligned}
    &\theta_{\I}(\omega) = \begin{cases}
        1, \;\;\omega\in (V_0+m,+\infty)\\
        0,\;\;\text{else}
    \end{cases},\;
    \theta_{\II}(\omega) = \begin{cases}
        1, \;\;\omega\in \left[V_0-m,V_0+m\right]\\
        0,\;\;\text{else}
    \end{cases},\\
    &\theta_{\III}(\omega) = \begin{cases}
        1, \;\;\omega\in (V_0+m,+\infty)\\
        0,\;\;\text{else}
    \end{cases},\;
    \theta_{\IV}(\omega) = \begin{cases}
        1, \;\;\omega\in \left[-m,m\right]\\
        0,\;\;\text{else}
    \end{cases},\\
    &\theta_{\V}(\omega) = \begin{cases}
        1, \;\;\omega\in (-\infty,-m)\\
        0,\;\;\text{else},
    \end{cases},\
    \;\;
    \chi(\omega) = \begin{cases}
        -1, \;\;\omega\in \left[m, V_0-m\right]\\
        1,\;\;\text{else}
    \end{cases}.
\end{aligned}
\eeq

Let us make several important remarks concerning the solutions (\ref{eq:in_modes_left})–(\ref{eq:in_modes_right}) and the notations (\ref{eq:notations}). First, in regions II and IV (where there are forbidden zones), the definitions (\ref{eq:notations}) imply $\pw=i|\pw|$ and $\kw=i|\kw|$, respectively, and in both cases $\varkappa_{\omega} = i|\varkappa_{\omega}|$ (in all other regions $\varkappa_{\omega}>0$). For this reason, the solutions (\ref{eq:in_modes_left}) and (\ref{eq:in_modes_right}) are valid in all frequency regions, including II and IV. In the latter two regions, incoming-from-the-right or incoming-from-the-left solutions are nonphysical. We therefore introduced the factors $(1-\theta_{\II})$ and $(1-\theta_{\IV})$, respectively, which explicitly set these nonphysical solutions to zero.

Second, the solutions (\ref{eq:in_modes_left}) and (\ref{eq:in_modes_right}) are chosen so that they can be interpreted as wave functions of an ``incoming'' electron in a single-particle interpretation. For this reason, the function $\chi(\omega)$ is introduced, which flips the direction of the momentum in region III (the ``Klein zone''). Then the directions of the probability current of the incident and transmitted waves are co-directed, and the transmission coefficient is positive (see also \cite{ternov2024paradoks15483} and Appendix \ref{app:1_particle_solution}).

For the mode in the Klein zone, the continuity of the spatial component of the current $j=\bar{\psi}\gamma^1\psi$ is defined as follows:
\begin{gather}
    \text{for} \quad \psiwR(x) : \quad j^{1}(x<0)= -1+\left(\frac{1-\kappaw}{1+\kappaw}\right)^2=-4 \frac{\kappaw}{\left(1+\kappaw\right)^2}=j^1(x>0),
\end{gather}
which expresses the conservation of probability. The phenomenon of Klein tunneling then occurs in the sense that when the energy is less than the height of the potential, there is a nonzero transmission probability, which does not occur for scalar particles in contrast to fermions. Moreover, in the infinite-potential limit, there is a nonvanishing transmission probability:
\begin{align}
    \lim_{V_0 \to \infty} T^2= 2\frac{\sqrt{\omega^2-m^2}}{\omega+\sqrt{\omega^2-m^2}}\ne0.
\end{align}

Third, the requirement that the modes (\ref{eq:in_modes_left})–(\ref{eq:in_modes_right}) are normalized to a delta function in energy
\beq\label{eq:norm_delta_function}
    \int dx\;\overleftarrow{\psi}_{\omega'}^{\dagger}\cdot\psiwL = \int dx\;\overrightarrow{\psi}_{\omega'}^{\dagger}\cdot\psiwR = 2\pi\delta\left(\w'-\w\right),
\eeq
leads to the normalization conditions for $A_{\omega}$ and $B_{\omega}$:
\beq\label{eq:norm_condition}
        \left|A_{\omega}\right|^2 = \frac{1}{2\pw\left|\omega-V_0-m\right|},\;\;\;
        \left|B_{\omega}\right|^2 = \frac{1}{2\kw\left|\omega-m\right|}.
\eeq
We fix arbitrary phases as
\beq\label{eq:norm_coefficients}
    A_{\omega} = \frac{1}{\sqrt{2\pw\left|\omega-V_0-m\right|}},\;\;B_{\omega}= \frac{\sign(m-\w)}{\sqrt{2\kw\left|\omega-m\right|}}.
\eeq
In this case the modes (\ref{eq:in_modes_left})--(\ref{eq:in_modes_right}) satisfy the simple relations
\beq\label{eq:modes_InterRelation}
\begin{aligned}
    \overleftarrow{\psi}_{-\omega+V_0}(x) = \gamma^0\gamma^1\overrightarrow{\psi}_{\omega}(-x), \;\;
    \overrightarrow{\psi}_{-\omega+V_0}(x) = \gamma^0\gamma^1\overleftarrow{\psi}_{\omega}(-x)\quad\text{for}\;\w\notin\left[m,V_0-m\right];
    \\
    \overleftarrow{\psi}_{-\omega+V_0}(x) = -\gamma^0\gamma^1\overrightarrow{\psi}_{\omega}^{*}(-x), \;\;
    \overrightarrow{\psi}_{-\omega+V_0}(x) = \gamma^0\gamma^1\overleftarrow{\psi}_{\omega}^{*}(-x)\quad\text{for}\;\w\in\left[m,V_0-m\right].
\end{aligned}
\eeq
Having at hand the orthonormal set of modes (\ref{eq:in_modes_left})--(\ref{eq:in_modes_right}) and the above remarks, we can expand the fermion field operator over the modes in the form: 

\begin{gather}\label{eq:decomposition_through_regions}
\widehat{\psi}(t,x) 
=\\=
\nonumber
\int_{m}^{V_0-m} \frac{d \omega}{2\pi}  \psiwR(x) e^{-i\omega t} \hataR_{\w}+\int_{V_0-m}^{V_0+m} \frac{d \omega}{2\pi}  \psiwR(x) e^{-i\omega t} \hataR_{\w}+\\+\nonumber \int_{V_0+m}^\infty \frac{d \omega}{2\pi}  \left(\psiwL(x) e^{-i\omega t} \hataL_{\w} + \psiwR(x) e^{-i\omega t} \hataR_{\w}\right)
+\\+ \nonumber
\int_{-(V_0-m)}^{-m} \frac{d \omega}{2\pi}  \overleftarrow{\psi}_{-\omega}(x) e^{i\omega t} \hatbL\phantom{\empty}^{\dagger}_{\w} 
+\int_{-m}^{m} \frac{d \omega}{2\pi} \overleftarrow{\psi}^*_{-\omega}(x) e^{i\omega t} \hatbL\phantom{\empty}^{\dagger}_{\w} + \\+\nonumber \int_m^\infty \frac{d \omega}{2\pi}  \left(\overleftarrow{\psi}^*_{-\omega}(x) e^{i\omega t} \hatbL\phantom{\empty}^{\dagger}_{\w} +\overrightarrow{\psi}^*_{-\omega}(x) e^{i\omega t} \hatbR\phantom{\empty}^{\dagger}_{\w} \right)
\end{gather}
or in a more compact form we have: 
\begin{gather}
\label{eq:fermion_decomposition_static}
    \widehat{\psi}(t,x) = \int_{-\infty}^{+\infty}\frac{d\omega}{2\pi}\Bigg\{\theta_{\I}(\w)\hataL_{\w}\psiwL(x)+ \bigg[\theta_{\I}(\w)+\theta_{\II}(\w)\bigg]\hataR_{\w}\psiwR(x)\; +  
    \\ \nonumber
    +\; 
    \theta_{\III}(\w)\bigg[\hataR_{\w}\psiwR(x) +   \hatbL\phantom{\empty}^{\dagger}_{\w}\psiwL(x)\bigg]+\bigg[\theta_{\IV}(\w)+\theta_{\V}(\w)\bigg]\hatbL\phantom{\empty}^{\dagger}_{-\w}\psiwL^*(x) + \theta_{\V}(\w)\hatbR\phantom{\empty}^{\dagger}_{-\w}\psiwR^*(x)\Bigg\}e^{-i\w t}\;\;.
\end{gather}
Here we introduced the operators $\hataL_{\w}, \hataR_{\w}, \hatbL_{\w}, \hatbR_{\w}$, which can be called annihilation operators of electron and positron states, depicted in Fig.\ref{Fig:In_modes_plot}. They satisfy the following anticommutation relations (all omitted anticommutators vanish):
\beq\label{eq:canonical_relations_opeators}
\begin{aligned}
     \left\{ \hataL_{\w}, \hataL\phantom{\empty}^{\dagger}_{\w'}  \right\} = 2\pi\delta\left(\w-\w'\right), \;
    \left\{ \hataR_{\w}, \hataR\phantom{\empty}^{\dagger}_{\w'}
    \right\} = 2\pi\delta\left(\w-\w'\right), 
    \\
    \left\{ \hatbL_{\w}, \hatbL\phantom{\empty}^{\dagger}_{\w'}  \right\} = 2\pi\delta\left(\w-\w'\right), \;
    \left\{ \hatbR_{\w}, \hatbR\phantom{\empty}^{\dagger}_{\w'}  \right\} = 2\pi\delta\left(\w-\w'\right),
\end{aligned}
\eeq
which, together with (\ref{eq:norm_condition}), guarantee the canonical anticommutation relations for the fermion field (see also \cite{Ochiai:2017slb}):
\beq\label{eq:canonical_relations_fields}
    \left\{ \widehat{\psi}_{a}(t,x), \widehat{\psi}\phantom{.}^{\dagger}_{b}(t,y)  \right\} = \delta_{ab}\cdot\delta\left(x-y\right),
\eeq
where $a,b$ denote the spinor indices.

\begin{figure}[t]
    \centering
    \includegraphics[width=0.8\linewidth]{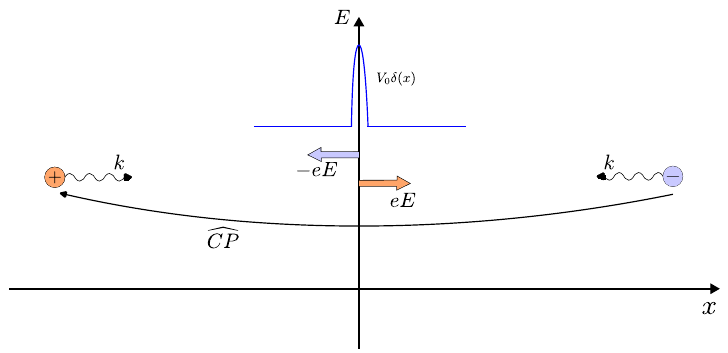}
    \caption{Illustration of the CP transformation.}
    \label{Fig:CP_illustration}
    \hfill
\end{figure}

Although the separation of the modes into ``electrons'' and ``positrons'' in the Klein zone is conventional, we split the operators in the way given in equations (\ref{eq:decomposition_through_regions})–(\ref{eq:fermion_decomposition_static}) for the following reason. The motion of an electron in an external electric field is equivalent to the motion of a positron in a mirror-reflected space in which the forces acting on the particles are inverted. At the level of solutions of the Dirac equation, one can introduce the charge-conjugation matrix $C = \gamma_0\gamma_1$ (with $\gamma^0=\gamma_0,\ \gamma^1=-\gamma_1$ in the metric that we use) and the parity matrix $P=i\gamma_0$. Hence, if a spinor $\Psi(t,x)$ solves the Dirac equation (\ref{eq:dirac}), then the $CP$-conjugated spinor $\Psi_{CP}(t,x) \equiv P C \overline{\Psi}^{,T}(t,-x) = i\gamma^0\gamma^1\Psi^{*}(t,-x)$ solves equation (\ref{eq:dirac}) with the opposite sign of the charge and the mirrored potential (as illustrated in Fig.\ref{Fig:CP_illustration}).
Interpreting the annihilation operators in (\ref{eq:fermion_decomposition_static}) as annihilators of the corresponding single-particle states allows one to define the $\widehat{CP}$ transformation on these operators as follows:
\beq\label{eq:CP_of_annihilation_operators}
    \widehat{CP}\phantom{\empty}^{\dagger}\;\hataL_{\omega}\;\widehat{CP} \equiv i\hatbR_{|-\omega+V_0|},\;\; \widehat{CP}\phantom{\empty}^{\dagger}\;\hataR_{\omega}\;\widehat{CP} \equiv i\hatbL_{|-\omega+V_0|}.
\eeq
Indeed, since $k_{-\omega+V_0} = p_{\omega}$ and vice versa, the transformation (\ref{eq:CP_of_annihilation_operators}) maps creation and annihilation operators of incoming particles into those of antiparticles (up to the standard phase factor $i$ of the $P$ transformation) incoming from the other side with the same momentum, as in Fig.\ref{Fig:CP_illustration}. Then, $CP$ conjugation of the field operator (\ref{eq:fermion_decomposition_static}) together with (\ref{eq:modes_InterRelation}) gives
\beq\label{eq:CP_of_fermion_field}
    \widehat{CP}\phantom{\empty}^{\dagger}\;\widehat{\psi}(t,x)\;\widehat{CP} = e^{-iV_0 t}\cdot\widehat{\psi}_{CP}(t,x),    
\eeq
where the phase can be removed by the shift $\omega\rightarrow \widetilde{\omega}=\omega-\frac{V_0}{2}$ in all equations. It can be regarded as a ``large gauge transformation'', which shifts the chemical potential but does not affect the expectation value of the current.

The fermion Hamiltonian in the step potential can be written in terms of creation and annihilation operators as
\begin{gather}
\label{eq:Hamiltonian}
    :\widehat{H}: \;= 
    \\ \nonumber
    = 
    \int_{0}^{\infty}\frac{d\w}{2\pi}\bigg\{\theta_{\I}(\w)\bigg[\w\bigg(\hataL\phantom{\empty}^{\dagger}_{\w}\hataL_{\w} + \hataR\phantom{\empty}^{\dagger}_{\w}\hataR_{\w}  \bigg)  + \Big(\w-V_0\Big)\bigg(\hatbL\phantom{\empty}^{\dagger}_{\w-V_0}\hatbL_{\w-V_0} + \hatbR\phantom{\empty}^{\dagger}_{\w-V_0}\hatbR_{\w-V_0}  \bigg)\bigg] +
    \\ \nonumber
    +\bigg(\theta_{\II}(\w)+\theta_{\III}(\w)\bigg)\bigg[\w \hataR\phantom{\empty}^{\dagger}_{\w}\hataR_{\w} + \Big(\w-V_0\Big)\hatbL\phantom{\empty}^{\dagger}_{\w-V_0}\hatbL_{\w-V_0} \bigg]\bigg\},
\end{gather}    
where we dropped an infinite constant in the normal ordering. From (\ref{eq:Hamiltonian}), one sees that the Fock vacuum $\Big|0\Big\rangle$ (by definition annihilated by the annihilation operators in (\ref{eq:fermion_decomposition_static})) is not the state of minimal eigenvalue of the Hamiltonian, which is bounded from below. It is easy to see that the ``ground state'' for our choice of mode basis is the state in which the frequencies in the interval $\w\in\left[0,V_0-m\right]$ are filled with positrons. We refer to it as the ground state $\Big|\text{gs}\Big\rangle$:
\beq\label{eq:ground_state_definition}
    \Big\langle\text{gs}\Big|                       \hatbL\phantom{\empty}^{\dagger}_{\w}\hatbL_{\w'}\Big|\text{gs}\Big\rangle = \theta(\w)\Big(1-\theta_{\I}(\w)-\theta_{\II}(\w)\Big)\cdot2\pi\delta\left(\w-\w'\right) \equiv \overleftarrow{n}^{(+)}_{\w} \cdot2\pi\delta\left(\w-\w'\right). 
\eeq
In many situations, even in the presence of strong external fields, there exists a state of the type (\ref{eq:ground_state_definition}) that realizes the minimum of the Hamiltonian—except for not fully physical situations, such as, e.g., an electric field that is constant in all space and eternal in time. In our problem, one can think of realizing such an initial state as $\Big|0\Big\rangle$ for the system under consideration being placed in a box of finite size. Then, the electron–positron pairs created by the electric field will gradually fill discrete energy levels until the electric current vanishes and the system reduces to the state $\Big|\text{gs}\Big\rangle$. (We will see below that the current is zero for the state $\Big|\text{gs}\Big\rangle$.) However, as the size of the box increases to infinity, the time required for the system to transition from $\Big|0\Big\rangle$ to the ground state $\Big|\text{gs}\Big\rangle$ tends to infinity. That is the reason why below we obtain a time-independent expression for the current.  

For a generic spatially homogeneous state $\Big|\text{st}\Big\rangle$ with nonzero occupation numbers, which is defined by the relations
\beq\label{eq:def_state_non_zero_occupation}
    \Big\langle \text{st}\Big|          \hatbL\phantom{\empty}^{\dagger}_{\w}\hatbL_{\w'}\Big|\text{st}\Big\rangle \equiv \overleftarrow{n}^{(+)}_{\w}\cdot2\pi\delta(\w-\w'),\;\;
     \Big\langle \text{st}\Big|          \hataR\phantom{\empty}^{\dagger}_{\w}\hataR_{\w'}\Big|\text{st}\Big\rangle \equiv \overrightarrow{n}^{(-)}_{\w}\cdot2\pi\delta(\w-\w')\;,
\eeq
one obtains the current of the form:
\beq\label{eq:current_eternal_field}
\begin{aligned}    \Big\langle\text{st}\Big|\widehat{j}^1(t,x)\Big|\text{st}\Big\rangle = \Big\langle\text{st}\Big| \widehat{\mathit{\Psi}}^{\dagger}(t,x)\gamma^0\gamma^1\widehat{\mathit{\Psi}}(t,x)\Big|\text{st}\Big\rangle = \int_{m}^{V_0-m}\frac{d\w}{2\pi}\Big[1-\overleftarrow{n}^{(+)}_{\w}- \overrightarrow{n}^{(-)}_{\w}\Big]\frac{-4\kappaw}{\left(\kappaw+1\right)^2}\;,
\end{aligned}
\eeq
where we assume that only the Klein zone and possibly regions $\II$ and $\IV$ are nontrivially populated. For the type of states that we consider here all zones except the Klein one do not contribute to the current due to a cancellation between left and right moving modes. In equation (\ref{eq:current_eternal_field}), the first term in the brackets appears when we commute the operators $\hatbL\phantom{\empty}^{\dagger}_{\w}$ and $\hatbL\phantom{\empty}_{\w}$ in the expression for $\hat{j}^1$. This term gives a nonvanishing contribution only in the Klein zone (region III) because left-moving modes in this region do not have right-moving counterparts (\ref{eq:fermion_decomposition_static}), which cancel the similar contributions in other regions.

In particular, for the state (\ref{eq:ground_state_definition}), one has 
\beq\label{eq:const_field_gs_current}
\boxed{
\Big\langle\text{gs}\Big|\widehat{j}^1(t,x)\Big|\text{gs}\Big\rangle \equiv 0\;,}
\eeq
as mentioned above. While for the Fock vacuum, i.e., the state without ``external particles'' with $\overleftarrow{n}^{(+)}_{\w} = \overrightarrow{n}^{(-)}_{\w} = 0$, we obtain the expression for the current known in the literature (see, e.g., the Appendix of \cite{Akhmedov:2020dgc})
\beq\label{eq:const_field_fock_vac_current}
\boxed{
\Big\langle0\Big|\widehat{j}^1(t,x)\Big|0\Big\rangle = \int_{m}^{V_0-m}\frac{d\w}{2\pi}\frac{-4\kappaw}{\left(\kappaw+1\right)^2}\;,}
\eeq
directed along the electric field from left to right, since the observable electric current equals $-e j^1$. Note that the current vanishes if $V_0 < 2m$.

One important point to emphasize is that the state $\Big|\text{gs}\Big\rangle$ can be realized as the Fock space ground state by quantizing the fermion field using a different basis of modes. For instance, instead of \eqref{eq:fermion_decomposition_static}, one can adopt the following decomposition:
\beq
\begin{aligned}
\label{eq:alternative_fermion_decomposition}
    \widehat{\psi}(t,x) = \int_{-\infty}^{+\infty}\frac{d\omega}{2\pi}\Bigg\{\theta(\w)\Big[\hatcL_{\w}\psiwL(x) + \hatcR_{\w}\psiwR(x)\Big]\; + 
    \\
    +\;\theta(-\w)\Big[\hatdL\phantom{\empty}^{\dagger}_{-\w}\psiwL^*(x) +\hatdR\phantom{\empty}^{\dagger}_{-\w}\psiwR^*(x))\Big]\Bigg\}e^{-i\w t}\;,
\end{aligned}
\eeq
where $\psiwL(x)$ and $\psiwR(x)$ are given by (\ref{eq:in_modes_left})–(\ref{eq:in_modes_right}). In this case, the Fock vacuum $\Big|0_1\Big\rangle$, which is annihilated by the operators $\widehat{c}$ and $\widehat{d}$, corresponds to the state of lowest energy and is therefore physically equivalent to $\Big|\text{gs}\Big\rangle$. In fact, the free Hamiltonian in such a case is given by
\beq\label{eq:free_Hamiltonian_alternative}
\begin{aligned}
    :\widehat{H}: \;=
    \int_{0}^{\infty}\frac{d\w}{2\pi}\cdot\w\cdot\bigg\{\hatcL\phantom{\empty}^{\dagger}_{\w}\hatcL_{\w}\Big(1-\theta_{\II}(\w)\Big) + \Big[\hatcR\phantom{\empty}^{\dagger}_{\w}\hatcR_{\w}  + \hatdR\phantom{\empty}^{\dagger}_{\w}\hatdR_{\w}\Big]\Big(1-\theta_{\IV}(\w)\Big)+\hatdL\phantom{\empty}^{\dagger}_{\w}\hatdL_{\w}  \bigg\}\;.
\end{aligned}
\eeq
In particular, the current is equal to zero for such a Fock space ground state:
\beq\label{eq:const_field_alternative_fock_current}
\boxed{
\Big\langle 0_1 \Big|\widehat{j}^1(t,x)\Big| 0_1 \Big\rangle \equiv 0\;.}
\eeq
To conclude this section, let us stress that in the situation of an external field that is constant in time, there is an ambiguity in the choice of what one interprets as particles and antiparticles. Furthermore, the current obviously depends on this interpretation and on the choice of the state. As we explained after Eq. (\ref{eq:ground_state_definition}), the current for any choice of the state is time independent, because one considers time independent background field and places the gaussian system in infinite volume.

Below, we continue with the consideration of situations in which there is no ambiguity in the definition of what is a particle and what is an antiparticle. Moreover, in those situations there will be a natural vacuum state before the background field is switched on.

\section{Switching on the electric field}\label{sec:Turning on the electric field}
In this section, we consider the situation in which there is no ambiguity in the choice and interpretation of the initial state. Let $-eA_{\mu}(t,x) = V_0\theta(t)\theta(x)\delta_{0\mu}$, i.e., the ``breakdown'' electric field is switched on at the moment $t=0$, and before switching on we assume that the theory was in the vacuum of empty Minkowski spacetime without external sources. We assume that the electric field is turned on almost instantaneously—on a timescale shorter than $\frac{1}{V_0}$. We clarify this point below.

We write the quantized fermion field with modes that solve the Dirac equation (\ref{eq:dirac}) with a time-dependent potential:
\beq\label{eq:decomposition_general}
\begin{aligned}
    \widehat{\mathit{\Psi}}(t,x) = \int_{-\infty}^{+\infty}\frac{d\omega}{2\pi}\bigg[\overleftarrow{\mathit{\Psi}}\phantom{\empty}^{(-)}_{\w}(t,x)\hataL_{\w} + \overrightarrow{\mathit{\Psi}}\phantom{\empty}^{(-)}_{\w}(t,x)\hataR_{\w} + \overleftarrow{\mathit{\Psi}}\phantom{\empty}^{(+)}_{\w}(t,x)\hatbL\phantom{\empty}^{\dagger}_{\w} + \overrightarrow{\mathit{\Psi}}\phantom{\empty}^{(+)}_{\w}(t,x)\hatbR\phantom{\empty}^{\dagger}_{\w}\bigg].
\end{aligned}
\eeq
Hence, before the switching on of the external field ($t<0$), the fermion field is quantized in the standard way, as in empty Minkowski spacetime:
\beq\label{eq:modes_before_electric_field}
\begin{aligned}
    \overleftarrow{\mathit{\Psi}}\phantom{\empty}^{(-)}_{\w}(t<0,x) = \overleftarrow{\psi}_{0}(\w,x)\theta(\w-m)e^{-i\w t},\quad\quad\overrightarrow{\mathit{\Psi}}\phantom{\empty}^{(-)}_{\w}(t<0,x) = \overrightarrow{\psi}_{0}(\w,x)\theta(\w-m)e^{-i\w t},
    \\
    \overleftarrow{\mathit{\Psi}}\phantom{\empty}^{(+)}_{\w}(t<0,x) = \overleftarrow{\psi}_{0}^{*}(\w,x)\theta(-\w-m)e^{-i\w t},\;\;\overrightarrow{\mathit{\Psi}}\phantom{\empty}^{(+)}_{\w}(t<0,x) = \overrightarrow{\psi}_{0}^{*}(\w,x)\theta(-\w-m)e^{-i\w t},
\end{aligned}
\eeq
where
\beq\label{eq:empty_space_modes}
    \overleftarrow{\psi}_{0}(\w,x) = \Big|B_{\w}\Big|
    \begin{pmatrix}
        -i\kw\\
        \w-m
    \end{pmatrix}e^{-i\kw x},\quad
     \overrightarrow{\psi}_{0}(\w,x) = B_{\w}
    \begin{pmatrix}
        i\kw\\
        \w-m
    \end{pmatrix}e^{i\kw x},
\eeq
and the functions $B_{\w}$, $\kw$ are defined in (\ref{eq:notations}), (\ref{eq:norm_coefficients}). 

\begin{figure}[t]
    \centering
    \includegraphics[width=1\linewidth]{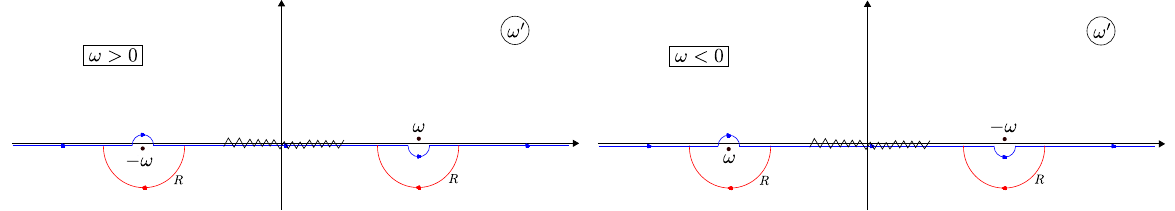}
    \caption{Contour of integration around the poles. The zigzag line indicates non-analyticity of the integrand elsewhere.}
    \label{Fig:contour}
    \hfill
\end{figure}

Choosing the vacuum $\Big|\text{vac}\Big\rangle$ as the initial state, which is annihilated by all annihilation operators $\widehat{a}$ and $\widehat{b}$, we find that in the asymptotic future $t\to+\infty$ only the ``positron'' solutions contribute to the current. For $t>0$, these solutions have the form
\beq\label{eq:solutions_after_electric_field}
\begin{aligned}
    \overleftarrow{\mathit{\Psi}}\phantom{\empty}^{(+)}_{\w}(t>0,x) = \int_{-\infty}^{+\infty}\frac{d\w'}{2\pi}\bigg[f_{11}\big(\w,\w'\big)\overleftarrow{\psi}_{\w'}(x) + f_{12}\big(\w,\w'\big)\overrightarrow{\psi}_{\w'}(x) \bigg]e^{-i\w't},
    \\
    \overrightarrow{\mathit{\Psi}}\phantom{\empty}^{(+)}_{\w}(t>0,x) = \int_{-\infty}^{+\infty}\frac{d\w'}{2\pi}\bigg[f_{21}\big(\w,\w'\big)\overleftarrow{\psi}_{\w'}(x) + f_{22}\big(\w,\w'\big)\overrightarrow{\psi}_{\w'}(x) \bigg]e^{-i\w't},
\end{aligned}
\eeq
where we expand the solution after the field is switched on over the basis (\ref{eq:in_modes_left})–(\ref{eq:in_modes_right}) with coefficient functions $f_{\bullet\bullet}(\w,\w')$, which are determined by matching, at $t=0$, the modes (\ref{eq:in_modes_left})–(\ref{eq:in_modes_right}) to the modes (\ref{eq:modes_before_electric_field}). For example,
\begin{gather}
\label{eq:example_f_coefficient}
    f_{11}(\w,\w') = \int_{-\infty}^{+\infty}dx\;\overleftarrow{\psi}\phantom{\empty}^{\dagger}_{\w'}(x)\cdot \overleftarrow{\psi}_{0}^{*}(\w,x) =
    \\ \nonumber
    =A_{\w'}\Big|B_{\w}\Big|\Bigg\{  
    \frac{2\kappawh^{*}\pwh^{*}\chi(\w')}{\kw'(\kappawh'+1)}\frac{-\kwh\kw+\Big(\w'-m\Big)\Big(\w-m\Big)}{\epsilon+i\Big(\kwh+\kw\Big)}- 
    \frac{\pwh\kw\chi(\w')-\Big(\w'-V_0-m\Big)\Big(\w-m\Big)}{\epsilon-i\Big(\pwh\chi(\w')+\kw\Big)} -
    \\ \nonumber
    -\frac{\kappawh^{*}-1}{\kappawh^{*}+1}\frac{\pwh\kw\chi(\w')+\Big(\w'-V_0-m\Big)\Big(\w-m\Big)}{\epsilon-i\Big(\pwh\chi(\w')-\kw\Big)}
    \Bigg\}\times\theta(-\w-m)\Big(1-\theta_{\II}(\w')\Big),
\end{gather}
with the prescription $\epsilon\rightarrow+0$. To find the behavior of the modes (\ref{eq:solutions_after_electric_field}) in the asymptotic future $t\rightarrow+\infty$ (more precisely, $t\gg\frac{1}{V_0}$), one cannot simply apply the Sokhotski–Plemelj theorem, keeping only the singular part, and at the same time invoke the Riemann–Lebesgue lemma. Indeed, in neighborhoods of the poles of the integral (\ref{eq:solutions_after_electric_field}) of size $\sim 1/t$, the integrand becomes large, $\sim t$, and is essentially non-oscillatory; therefore, the finite parts of the coefficients $f_{\bullet\bullet}(\w,\w')$ also contribute in the large-time limit, and to extract the leading terms one must proceed more carefully. We isolate poles in the integration variable $\w'$ using an identity of the form:
\beq\label{eq:omega_poles}
    \kwh-\kw-i\epsilon = \frac{1}{\kwh+\kw}\Big(\w'-\w-i\epsilon\cdot\sign(\w'+\w)\Big)\Big(\w'+\w-i\epsilon\cdot\sign(\w'-\w)\Big).
\eeq
Then we use that in a small neighborhood of each pole the integrand in (\ref{eq:solutions_after_electric_field}) is analytic and add in the lower half-plane $\text{Im}\left\{\w'\right\}<0$ a semicircle of radius $R$ (see Fig.\ref{Fig:contour}; one can choose e.g. $R\sim \frac{V_0}{\sqrt{V_0t}}\ll V_0$). In all regions of integration in (\ref{eq:solutions_after_electric_field}) except neighborhoods of radius $R$ around the poles, the contributions vanish by the Riemann–Lebesgue lemma as $t\rightarrow\infty$, whereas in the vicinity of the poles, we can apply Cauchy's theorem due to the analyticity of the integrand. Furthermore, the integrals over the red semicircles in Fig.\ref{Fig:contour} also tend to zero at large times, as they lie in the lower half-plane.
Proceeding in this way, we obtain the expressions for the modes in the asymptotic future:
\begin{gather}
\label{eq:left_mode_asymptotic_future}
    \overleftarrow{\mathit{\Psi}}\phantom{\empty}^{(+)}_{\w}(t\rightarrow\infty,x) \approx
    \\ 
    \nonumber
    \approx\;\theta(-\w-V_0-m)\cdot\Big|B_{\w}\Big|B_{\w+V_0}\frac{-4k_{\w+V_0}\Big(\w-m\Big)}{\varkappa_{\w+V_0}+1}\overrightarrow{\psi}_{\omega+V_0}(x)e^{-i(\w+V_0)t}\;+\;
    \\ 
    \nonumber
    +\;\theta(-\w-m)\bigg[\frac{1-\chi(\w+V_0)}{2}-\frac{1+\chi(\w+V_0)}{2}\frac{\varkappa_{\w+V_0}^{*}-1}{\varkappa_{\w+V_0}^{*}+1}\bigg]\overleftarrow{\psi}_{\omega+V_0}(x)e^{-i(\w+V_0)t} \; ;
\end{gather}
and
\begin{gather}
\label{eq:right_mode_asymptotic_future}
    \overrightarrow{\mathit{\Psi}}\phantom{\empty}^{(+)}_{\w}(t\rightarrow\infty,x) \approx 
    \\ \nonumber
    \approx\;A_{\w}B_{\w}\frac{4\kappaw\pw\Big(m-\w\Big)}{\kappaw+1}\cdot\theta(-\w-m)\psiwL(x)e^{-i\w t} + \frac{\kappaw-1}{\kappaw+1}\theta(-\w-m)\psiwR(x)e^{-i\w t}.
\end{gather}

Note that at large frequencies $|\w|\gg V_0$ the modes (\ref{eq:left_mode_asymptotic_future})–(\ref{eq:right_mode_asymptotic_future}) coincide to good accuracy with plane waves. At the same time, only modes with $\w<V_0$ contribute to the current. In particular, this implies that our results for instantaneous switching on of the field work with good accuracy if the actual switching time satisfies $\delta t \ll \frac{1}{V_0}$.

Using (\ref{eq:left_mode_asymptotic_future})–(\ref{eq:right_mode_asymptotic_future}), we find the current in the asymptotic future:
\begin{gather}
    \Big\langle\text{vac}\Big| \widehat{\mathit{\Psi}}^{\dagger}(t,x)\gamma^0\gamma^1\widehat{\mathit{\Psi}}(t,x)\Big|\text{vac}\Big\rangle\bigg|_{t\rightarrow\infty} =
    \\ \nonumber
    =\int_{-\infty}^{+\infty}\frac{d\w}{2\pi}\bigg[\overleftarrow{\mathit{\Psi}}\phantom{\empty}^{(+)\dagger}_{\w}(t,x) \gamma^0\gamma^1 \overleftarrow{\mathit{\Psi}}\phantom{\empty}^{(+)}_{\w}(t,x) + \overrightarrow{\mathit{\Psi}}\phantom{\empty}^{(+)\dagger}_{\w}(t,x) \gamma^0\gamma^1 \overrightarrow{\mathit{\Psi}}\phantom{\empty}^{(+)}_{\w}(t,x)\bigg]\Bigg|_{t\rightarrow\infty}, \end{gather}
which is given by: 
\begin{align}
\label{eq:current_asymptotic_future}
    \boxed{\Big\langle\text{vac}\Big|j^1(t,x)\Big|\text{vac}\Big\rangle\bigg|_{t\rightarrow\infty} \simeq \;
    \int_{m}^{V_0-m}\frac{d\w}{2\pi}\frac{-4\kappaw}{\left(\kappaw+1\right)^2}\;\;. }
\end{align}
Thus, in this case the current coincides with the current in the state $\Big|0\Big\rangle$ from the previous section. Let us stress again that, as we explained after Eq. (\ref{eq:ground_state_definition}), the obtained current is time independent, because one considers the system in infinite volume. It is probably worth stressing that to observe the relaxation from the initial state to the state with zero current one has to consider interacting theory in 4D rather that the gaussian one in 2D.

\section{Current for a generic potential of the form $U(x)\theta(t)$}
\label{Lapushkin}

The main purpose of this section is to calculate the current for a generic situation. The initial state is the vacuum (the Poincaré-invariant state with the lowest energy in empty Minkowski spacetime), and at $t=0$ we turn on an external potential $U(x)$. We assume that $\lim\limits_{x\rightarrow-\infty} U(x) = 0$ and $\lim\limits_{x\rightarrow+\infty} U(x) = V_0 > 2m$, and we assume the same conditions as in the previous section. Thus, for $t<0$ the basis of modes is as in (\ref{eq:empty_space_modes}) or (\ref{left free}), (\ref{right free}):

\begin{gather}\label{Lap1}
\mathit{\Psi}^{in}_{\omega,\sigma}(t,x)= e^{-i\omega t} \psi_{0,\sigma}(\omega, x) = \frac{e^{-i\omega t}}{\sqrt{2k_\omega |\omega-m|}}e^{i k_\omega\sigma x}\begin{pmatrix}
		ik_\omega\sigma\\
		\omega-m
	\end{pmatrix}, \,\, k_\omega = \sqrt{\omega^2-m^2}, \,\, \sigma=\pm1.
\end{gather}
We simply use different notations here.
The reason for this is as follows. In the previous sections, part of the Hamiltonian spectrum was doubly degenerate, and we distinguished the modes corresponding to one degenerate energy level by arrows above the modes. However, for a generic potential, right- and left-moving waves are not orthogonal to each other. Hence, to keep an orthonormal basis of modes, we will denote and distinguish them by other means.

We want to find the behavior of these modes as $t \to +\infty$, i.e., after the switching on of the potential $U(x)$. This behavior will be used to find the current at future infinity. Essentially, we have to perform here the same type of calculations as in the previous section. We now define a convenient basis of exact harmonics in the potential $U(x)$ to decompose the in-modes, $\mathit{\Psi}^{in}_{\omega,\sigma}(t,x)$, as $t \to +\infty$. Namely, the basis consists of the same three types (for different regions of values of $\omega$) of modes as in the previous sections, but we use the index $\sigma$ instead of arrows to designate the modes, as mentioned above: we write $\mathit{\Psi}^{(0)}_{\omega,\sigma},\sigma=\pm1$ instead of $\overrightarrow{\mathit{\Psi}}_{\omega}$ or $\overleftarrow{\mathit{\Psi}}_{\omega}$. 

Thus, for $\omega\in(-\infty;-m]\cup[m;V_0-m]\cup[V_0+m;+\infty)$ we denote the modes in the basis under discussion as $\mathit{\Psi}^{(0)}_{\omega,\sigma}(x,t) = e^{-i\omega t} \psi^{(0)}_{\omega, \sigma}(x)$. For $\omega\in[-m,m]$ we have modes that exponentially decay as $x\to -\infty$ (similarly to the situation depicted in Fig.\ref{Fig:In_modes_plot}). We denote such modes as $\mathit{\Psi}^{(0)}_{\omega,\leftarrow}(x) = e^{-i\omega t} \psi^{(0)}_{\omega, \leftarrow}(x)$. Similarly, in the region $\omega\in[V_0-m;V_0+m]$ the modes are denoted as $\mathit{\Psi}^{(0)}_{\omega,\rightarrow}(x) = e^{-i\omega t} \psi^{(0)}_{\omega, \rightarrow}(x)$. They decay exponentially as $x\to +\infty$.

Furthermore, to estimate the current, we need to use only the asymptotic form of the modes $\mathit{\psi}^{(0)}_{\omega,\sigma}(x)$ as $|x|\rightarrow\infty$, where one essentially has to solve the free Dirac equation; we assume that the solution of the scattering problem in the potential $U(x)$ is known. 
The current will be expressed in terms of the scattering data.
The asymptotic form of  $\mathit{\psi}^{(0)}_{\omega,\leftarrow}(x)$ as $x\rightarrow -\infty$ is as follows:
\begin{gather}
\mathit{\psi}^{(0)}_{\omega,\leftarrow}(x)\approx\theta(m-|\omega|)K_{-}(\omega)e^{\sqrt{m^2-\omega^2}x}\begin{pmatrix}
		\sqrt{m^2-\omega^2}\\
		\omega-m
	\end{pmatrix}.
\end{gather}
At the same time the asymptotic form of $\mathit{\psi}^{(0)}_{\omega,\rightarrow}(x)$ for $x\rightarrow -\infty$ is:
\begin{gather}
\mathit{\psi}^{(0)}_{\omega,\rightarrow}(x)\approx\theta\Big(m-|\omega-V_0|\Big)\sum_{\sigma'}K_{-}(\sigma';\omega)e^{ik_\omega\sigma'x}\begin{pmatrix}
		ik_\omega\sigma'\\
		\omega-m
	\end{pmatrix}.
\end{gather}
Furthermore, for $x\rightarrow +\infty$ these modes behave as ($p_\omega = \sqrt{(\omega-V_0)^2-m^2}$):
\begin{gather}
\mathit{\psi}^{(0)}_{\omega,\leftarrow}(x)\approx\theta(m-|\omega|)\sum_{\sigma'}K_{+}(\sigma';\omega)e^{ip_\omega\sigma'x}\begin{pmatrix}
		ip_\omega\sigma'\\
		\omega-V_0-m
	\end{pmatrix},
\end{gather}
and 
\begin{gather}
\mathit{\psi}^{(0)}_{\omega,\rightarrow}(x)\approx\theta\Big(m-|\omega-V_0|\Big)K_{+}(\omega)e^{-\sqrt{m^2-(\omega-V_0)^2}x}\begin{pmatrix}
		-\sqrt{m^2-(\omega-V_0)^2}\\
		\omega-V_0-m
	\end{pmatrix}.
\end{gather}
The coefficients $K_{\pm}$ in the asymptotic form of the modes can be found from the single-particle problem in the potential $U(x)$ for each mode separately, but these coefficients will not contribute to the current. The contribution to the current will apparently come from the Klein zone, $\omega \in [m, V_0 - m]$.

In the regions $\omega\in(-\infty;m]\cup[m;V_0-m]\cup[V_0+m;+\infty)$ the modes behave as:
\begin{gather} \label{asym1}
\mathit{\psi}^{(0)}_{\omega,\sigma}(x)\approx\theta(|\omega|-m)\theta\Big(|\omega-V_0|-m\Big)\sum_{\sigma'}\frac{N(\sigma,\sigma';\omega)}{\sqrt{2k_\omega|\omega-m|}}e^{ik_\omega\sigma'x}\begin{pmatrix}
		ik_\omega\sigma'\\
		\omega-m
	\end{pmatrix},
\end{gather}
as $x\rightarrow-\infty$.
At the same time, for $x\rightarrow +\infty$ these modes behave as:
\begin{gather}\label{asym2}
\mathit{\psi}^{(0)}_{\omega,\sigma}(x)\approx \theta(|\omega|-m)\theta\Big(|\omega-V_0|-m\Big) \sum_{\sigma',\sigma''}\frac{N(\sigma,\sigma'';\omega)}{\sqrt{2p_\omega|\omega-V_0-m|}} S(\sigma'',\sigma';\omega)e^{ip_\omega\sigma'x}\begin{pmatrix}
		ip_\omega\sigma'\\
		\omega-V_0-m
	\end{pmatrix},
\end{gather}
where $S$ is the transfer matrix. Its elements can be found by solving the scattering problem in the potential $U(x)$, and $N$ is some matrix, which we calculate below using the current conservation law and normalization conditions.



Thus, the expansion of $\mathit{\Psi}^{in}_{\omega,\sigma}(t,x)$ for $t > 0$ over the basis of the modes $\mathit{\psi}^{(0)}_{\omega,\sigma}$ is as follows:
\begin{gather}
\mathit{\Psi}^{in}_{\omega,\sigma}(t,x)=\sum_{\sigma'}\int^{-m}_{-\infty}f_{\sigma,\sigma'}(\omega,\omega')\mathit{\psi}^{(0)}_{\omega',\sigma'}(x)e^{-i\omega't}\frac{d\omega'}{2\pi}+\sum_{\sigma'}\int^{V_0-m}_{m}f_{\sigma,\sigma'}(\omega,\omega')\mathit{\psi}^{(0)}_{\omega',\sigma'}(x)e^{-i\omega't}\frac{d\omega'}{2\pi}+\nonumber\\+\sum_{\sigma'}\int^{+\infty}_{V_0+m}f_{\sigma,\sigma'}(\omega,\omega')\mathit{\psi}^{(0)}_{\omega',\sigma'}(x)e^{-i\omega't}\frac{d\omega'}{2\pi}+\int^{m}_{-m}f_{\sigma,\leftarrow}(\omega,\omega')\mathit{\psi}^{(0)}_{\omega',\leftarrow}(x)e^{-i\omega't}\frac{d\omega'}{2\pi}+\nonumber\\+\int^{V_0+m}_{V_0-m}f_{\sigma,\rightarrow}(\omega,\omega')\mathit{\psi}^{(0)}_{\omega',\rightarrow}(x)e^{-i\omega't}\frac{d\omega'}{2\pi}. \label{Lap2}
\end{gather}
The kernels $f$ are similar to those introduced in the previous section. They are the expansion coefficients of the function $\mathit{\psi}_{0,\sigma}(\omega, x)$ from (\ref{Lap1}) over the basis of functions $\mathit{\psi}^{(0)}_{\omega,\sigma}(x)$:
\begin{gather}
f_{\sigma,\sigma'}(\omega,\omega')=\int^{+\infty}_{-\infty}\mathit{\psi}^{(0)*}_{\omega',\sigma'}(x)\mathit{\psi}_{0,\sigma}(\omega,x)dx, \quad {\rm and} \\
f_{\sigma,\leftarrow}(\omega,\omega')=\int^{+\infty}_{-\infty}\mathit{\psi}^{(0)*}_{\omega',\leftarrow}(x)\mathit{\psi}_{0,\sigma}(\omega,x)dx, \quad 
f_{\sigma,\rightarrow}(\omega,\omega')=\int^{+\infty}_{-\infty}\mathit{\psi}^{(0)*}_{\omega',\rightarrow}(x)\mathit{\psi}_{0,\sigma}(\omega,x)dx. \nonumber
\end{gather}
Since $U(x)$ is a measurable and bounded function, we can approximate it by finite step functions. Using this fact, we can prove that ${\rm sup}_{x\in[-a,a]}|(\mathit{\psi}^{in*}_{0,\sigma'}(\omega',x)\mathit{\psi}^{(0)}_{\omega,\sigma}(x))|$ is a finite function of $\omega$ and $\omega'$ for any fixed $a>0$. In the equations below, we can choose any $a>0$. The answer for the current will not depend on the choice of $a$. In fact, one can make the following decomposition:
\begin{gather}
f_{\sigma,\sigma'}(\omega,\omega')=\int^{+\infty}_{-\infty} \mathit{\Psi}^{in*}_{\omega',\sigma'}(0,x)\mathit{\Psi}^{(0)}_{\omega,\sigma}(x) \, dx = \int^{-a}_{-\infty} \mathit{\Psi}^{in*}_{0,\sigma'}(\omega',x)\mathit{\Psi}^{(0)}_{\omega,\sigma}(x) \, dx + \nonumber \\ + \int^{+\infty}_a \mathit{\Psi}^{in*}_{0,\sigma'}(\omega',x)\mathit{\Psi}^{(0)}_{\omega,\sigma}(x) \, dx + {\rm reg} \,  f^{(a)}_{\sigma,\sigma'}(\omega,\omega') = {\rm sing}\, f^{(a)}_{\sigma,\sigma'}(\omega,\omega') + {\rm reg}\, f^{(a)}_{\sigma,\sigma'}(\omega,\omega'), \label{Decompeq}
\end{gather}
where reg $f^{(a)}_{\sigma,\sigma'}(\omega,\omega')$ is bounded and continuous function of arguments $\omega$ and $\omega'$. Similarly one can define reg $f^{(a)}_{\sigma,\leftarrow}(\omega,\omega')$ and reg $f^{(a)}_{\sigma,\rightarrow}(\omega,\omega')$. 

For simplicity, we will analyze only the first integral in Eq. (\ref{Lap2}). The analysis of the other integrals is similar. Using Riemann's oscillatory theorem, we can prove that the integral with $\text{reg}\;f^{(a)}_{\sigma,\sigma'}(\omega,\omega')$ is equal to zero in the limit $t\rightarrow +\infty$, and:

\begin{gather}
\sum_{\sigma'}\int^{-m}_{-\infty}f_{\sigma,\sigma'}(\omega,\omega')\mathit{\psi}^{(0)}_{\omega',\sigma'}(x)e^{-i\omega't}\frac{d\omega'}{2\pi}=\sum_{\sigma'}\int^{-m}_{-\infty} {\rm sing} \, f^{(a)}_{\sigma,\sigma'}(\omega,\omega')\mathit{\psi}^{(0)}_{\omega',\sigma'}(x)e^{-i\omega't}\frac{d\omega'}{2\pi}+\nonumber\\+\sum_{\sigma'}\int^{-m}_{-\infty}{\rm reg} \, f^{(a)}_{\sigma,\sigma'}(\omega,\omega')\mathit{\psi}^{(0)}_{\omega',\sigma'}(x)e^{-i\omega't}\frac{d\omega'}{2\pi}
\approx \sum_{\sigma'}\int^{-m}_{-\infty}{\rm sing} \, f^{(a)}_{\sigma,\sigma'}(\omega,\omega')\mathit{\psi}^{(0)}_{\omega',\sigma'}(x)e^{-i\omega't}\frac{d\omega'}{2\pi}.
\end{gather}
Thus, we need to calculate only $\text{sing}\; f^{(a)}_{\sigma,\sigma'}(\omega,\omega')$ for some $a$. We can choose $a$ large enough so that the asymptotic form of the modes (\ref{asym1}), (\ref{asym2}) can be used in the region $|x|>a$. Then, from (\ref{Decompeq}) we obtain the following coefficient function:

\begin{gather}
{\rm sing} \, f^{(a)}(\omega,\sigma;\omega',\sigma')\simeq \frac{i\sigma|\omega|}{k_\omega}\left(N^*(\sigma',\sigma;\omega)\frac{\theta(\omega \omega')^{\phantom{\frac12}}}{k_{\omega'} - k_\omega + i0\sigma} -\right. \nonumber \\ \left. -\sum_{\sigma''}N^*(\sigma',\sigma'';\omega+V)S^*(\sigma'',\sigma;\omega+V)\frac{\theta\Big[(\omega+V_0)\omega'\Big]}{p_{\omega'}-k_\omega -i0\sigma}\right).
\end{gather}
As one can see, this answer does not depend on $a$. The rest of the calculation is similar to that in the previous section, and the result for (\ref{Lap2}) is:
\begin{gather}
\mathit{\Psi}^{in}_{\omega,\sigma}(t,x)\approx\delta_{\sigma,sgn(\omega)}e^{-i\omega t}\Big[\sqrt{2 k _\omega |\omega-m|}K^*_{-}(\sigma;\omega)\mathit{\Psi}^{(0)}_{\omega,\rightarrow}(x)+\sum_{\sigma'}N^*(\sigma',\sigma;\omega)\mathit{\Psi}^{(0)}_{\omega,\sigma'}(x)\Big]+\nonumber\\
+\delta_{\sigma,-sgn(\omega)}e^{-i(\omega+V_0)t}\Big[\sqrt{2 k_\omega |\omega-m|} K^*_{+}(\sigma;\omega+V_0)\mathit{\Psi}^{(0)}_{\omega+V_0,\leftarrow}(x) + \nonumber \\ + \sum_{\sigma',\sigma''}N^*(\sigma',\sigma'';\omega+V_0)S^*(\sigma'',\sigma;\omega+V_0)\mathit{\Psi}^{(0)}_{\omega+V_0,\sigma'}(x)\Big], \quad {\rm as} \quad t \to +\infty.
\end{gather}
To proceed further, we have to find the matrices $N$ and $S$. It is straightforward to prove that

$$\mathit{\psi}^{(0)*}_{\omega,\sigma}(x)\hat{\alpha}\mathit{\psi}^{(0)}_{\omega,\sigma'}(x) = {\rm const}$$
and 

$$
\mathit{\psi}^{(0)*}_{\omega,\leftarrow}(x) \hat{\alpha} \mathit{\psi}^{(0)}_{\omega,\leftarrow}(x) = \mathit{\psi}^{(0)*}_{\omega,\rightarrow}(x) \hat{\alpha} \mathit{\psi}^{(0)}_{\omega,\rightarrow}(x) = 0,$$ 
where $\hat{\alpha}$ is the Dirac $\alpha$ matrix in 2D.
From these equations, we obtain restrictions on the matrices $N$ and $S$:
\begin{gather}
\psi^{(0)*}_{\omega,\sigma}(x)\hat{\alpha}\psi^{(0)}_{\omega,\sigma'}(x)|_{x\rightarrow-\infty} = {\rm sign} \, (\omega-V_0-m)\sum_{\sigma''}S^*(\sigma_1,\sigma'';\omega)\sigma''S(\sigma_2,\sigma'';\omega)=\nonumber\\ = {\rm sign} \, (\omega-m)\sigma_1\delta_{\sigma_1,\sigma_2}= \psi^{(0)*}_{\omega,\sigma}(x)\hat{\alpha}\psi^{(0)}_{\omega,\sigma'}(x)|_{x\rightarrow+\infty}.
\end{gather}
Other restrictions follow from the normalization condition:

$$
\int^{+\infty}_{-\infty}\mathit{\Psi}^{(0)*}_{\omega,\sigma}(x)\mathit{\Psi}^{(0)}_{\omega',\sigma'}(x)dx = \delta_{\sigma\sigma'} \, \delta(\omega - \omega').
$$
We can introduce the same variable $a$ as above:
\begin{gather}
\delta_{\sigma\sigma'}\delta(\omega - \omega')=\int^{-a}_{-\infty}\mathit{\Psi}^{(0)*}_{\omega,\sigma}(x)\mathit{\Psi}^{(0)}_{\omega',\sigma'}(x)dx +\int^{a}_{-a}\mathit{\Psi}^{(0)*}_{\omega,\sigma}(x)\mathit{\Psi}^{(0)}_{\omega',\sigma'}(x)dx + \int^{+\infty}_{a}\mathit{\Psi}^{(0)*}_{\omega,\sigma}(x)\mathit{\Psi}^{(0)}_{\omega',\sigma'}(x)dx
\end{gather}
Again, the integral over the interval $[-a;a]$ will be a bounded function of its arguments. Therefore, the delta function can appear only from the other two integrals, which can be calculated using the equations (\ref{asym1}) and (\ref{asym2}):
\begin{gather}
\delta_{\sigma\sigma'}\delta(\omega - \omega')=\frac{1}{2}\sum_{\sigma_1,\sigma_2}N^*(\sigma,\sigma_1;\omega)\left[\delta_{\sigma_1,\sigma_2}+\sum_{\sigma''}S^*(\sigma_1,\sigma'')S(\sigma_2,\sigma'')\right]N(\sigma',\sigma_2;\omega) \delta(\omega - \omega')+\nonumber\\+{\rm reg}_{\sigma,\sigma'}(\omega,\omega'),
\end{gather}
where ${\rm reg}_{\sigma,\sigma'}(\omega,\omega')$ is some regular function of its arguments. To satisfy the normalization condition\footnote{A similar procedure has been used in \cite{Akhmedov:2019rvx}, \cite{Lanina:2020yvh}.} the regular term should be set to $0$ by a choice of the matrix $N$, and we obtain that:
\begin{gather}\label{normcond}
\sum_{\sigma_1,\sigma_2}N^*(\sigma,\sigma_1;\omega)\left[\delta_{\sigma_1,\sigma_2}+\sum_{\sigma''}S^*(\sigma_1,\sigma'')S(\sigma_2,\sigma'')\right]N(\sigma',\sigma_2;\omega) = 2\, \delta_{\sigma,\sigma'}.
\end{gather}
Again, $a$ drops out from the final relation. Now we can rewrite and solve these conditions using the components of the $S$ matrix:
\begin{gather}\label{matrxiel}
\hat{S}(\omega)=\begin{pmatrix}
		S(1,1;\omega) & S(1,-1;\omega)\\
		S(-1,1;\omega) & S(-1,-1;\omega)
	\end{pmatrix}=\begin{pmatrix}
		\alpha(\omega) & \beta(\omega)\\
		\beta^*(\omega) & \alpha^*(\omega)
	\end{pmatrix},
\end{gather}
where, from the current conservation law, we obtain that 

$$
|\alpha(\omega)|^2-|\beta(\omega)|^2={\rm sign} \, \left(\frac{\omega-m}{\omega-V_0-m}\right).
$$ 
Furthermore, from the normalization condition (\ref{normcond}) we obtain that: 
\begin{gather}
\hat{N}^{\dagger}(\omega)\hat{N}(\omega)=\left(\frac{1+\hat{S}^{\dagger}(\omega)\hat{S}(\omega)}{2}\right)^{-1}
\end{gather}
Now we are ready to calculate the current. The field operator is:
\begin{gather}
\hat{\mathit{\Psi}}(t,x)=\sum_{\sigma}\int^{+\infty}_{m} \left[\mathit{\Psi}^{in}_{\omega,\sigma}(t,x)\hat{a}^{-}_{\omega,\sigma}+\mathit{\Psi}^{in}_{-\omega,\sigma}(t,x)\hat{b}^{+}_{\omega,\sigma}\right]\frac{d\omega}{2\pi}.
\end{gather}
Then the current is given by:
\begin{gather}
\Big\langle0\Big|\hat{j}^{1}(t,x)\Big|0\Big\rangle\Big|_{t\rightarrow+\infty}= \sum_\sigma \int_{m}^{+\infty}\mathit{\Psi}^{in*}_{-\omega,\sigma}(t,x)\hat{\alpha}\mathit{\Psi}^{in}_{-\omega,\sigma}(t,x)\frac{d\omega}{2\pi}\Big|_{t\rightarrow+\infty}=\nonumber\\=-\int^{+\infty}_{m}\Bigg\{\Big[\hat{S}^{\dagger}(-\omega)\hat{N}^{\dagger}(-\omega)\hat{N}(-\omega)\hat{\sigma}_z\hat{N}^{\dagger}(-\omega)\hat{N}(-\omega)\hat{S}(-\omega)\Big]_{1,1}+ \\ \nonumber+\Big[\hat{N}^{\dagger}(-\omega)\hat{N}(-\omega)\hat{\sigma}_z\hat{N}^{\dagger}(-\omega)\hat{N}(-\omega)\Big]_{-1,-1}\Bigg\}\frac{d\omega}{2\pi}+\\\nonumber+\theta(V_0-2m)\int^{V_0-m}_{m}\Big[\hat{S}^{\dagger}(\omega)\hat{N}^{\dagger}(\omega)\hat{N}(\omega)\hat{\sigma}_z\hat{N}^{\dagger}(\omega)\hat{N}(\omega)\hat{S}(\omega)\Big]_{1,1}\frac{d\omega}{2\pi},
\end{gather}
where the brackets with indices below denote the corresponding element of this matrix in the notation of (\ref{matrxiel}).
Finally, we obtain that:
\begin{gather}
\boxed{\left\langle0\left|\hat{j}^{1}(t,x)\right|0\right\rangle\Big|_{t\to +\infty}\simeq -\int^{V_0-m}_{m}\frac{1}{|\beta(\omega)|^2}\frac{d\omega}{2\pi}.}
\end{gather}
\
Thus, the current can be expressed in terms of the scattering coefficients in the potential $U(x)$ and is nonzero for $V_0> 2m$.

\section{Finite duration of the external field}\label{sec:Finite action of the external electric field} 
Above, we have computed particle currents for cases in which the external field is either always on or switched on at some moment in time. In this section, we consider a situation in which the external field is switched on at some time $t = -T$ and then switched off at $t = T$. A direct calculation of the expectation value of the current in such a situation is very complicated. However, in this case we have well-defined $In$ and $Out$ Fock space ground states and can therefore compute the number of created particles.

The electromagnetic potential is given by:
\begin{align}
    A_0 = V \theta(x) \theta(T^2 - t^2) \quad \text{and} \quad A_1 = 0.
\end{align}
We can represent solutions of the Dirac equation in such a potential in the form:
\begin{align}
\label{solution}
    \psi = \theta(-T - t) \psi_{t<-T} + \theta(T^2 - t^2) \psi_{-T < t < T} + \theta(t - T) \psi_{t>T},
\end{align}
where $\psi_{t<-T}$ and $\psi_{t>T}$ are solutions of the free Dirac equation, which can be expanded in terms of $(u_i,v_i)$—the set of left- and right-moving positive- and negative-energy modes. The index $i$ is used to shorten the notation below. The positive-energy modes are:
\begin{align}
\label{not1}
    u_i^t = e^{-i \omega_i^u t} f^u_i = ( \overleftarrow{u}_i, \overrightarrow{u}_i),
\end{align}
where:
\begin{align}
    \overleftarrow{u}^t_i = e^{-i \omega^u_i t} \overleftarrow{\psi}_{\omega^u_i} \quad \text{and} \quad \overrightarrow{u}^t_i = e^{-i \omega^u_i t} \overrightarrow{\psi}_{\omega^u_i},
\end{align}
and $\psi_{\omega^u_i}^L$ and $\psi_{\omega^u_i}^R$ are defined in \eqref{left free} and \eqref{right free}. The set $v_i$ consists of left- and right-moving negative-energy modes:
\begin{align}
\label{not2}
    v^t_i = e^{i \omega_i^v t} g^v_i = ( \overleftarrow{v}_i, \overrightarrow{v}_i),
\end{align}
where
\begin{align}
    \overleftarrow{v}^t_i = e^{i \omega^v_i t} \overleftarrow{\psi}^*_{-\omega^v_i} \quad \text{and} \quad \overrightarrow{v}^t_i = e^{i \omega^v_i t} \overrightarrow{\psi}^*_{-\omega^v_i}.
\end{align}
The modes obey the orthonormality conditions:
\begin{align}
    \langle u_i, u_j \rangle = \delta_{ij}, \quad 
    \langle v_i, v_j \rangle = \delta_{ij}, \quad \text{and} \quad 
    \langle v_i, u_j \rangle = 0,
\end{align}
and the completeness relation:
\begin{align}
\sum_i |u_i\rangle\langle u_i| + \sum_i |v_i\rangle\langle v_i| = \hat{1}.
\end{align}

In \eqref{solution}, the solution of the Dirac equation with the external electric field is denoted by $\psi_{-T < t < T}$. It can be expanded in terms of the complete and orthonormal set of positive- and negative-frequency solutions $(U_i, V_i)$. The positive-energy set is:
\begin{align}
\label{not3}
    U^t_i = e^{-i \omega_i^U t} F^U_i = ( \overleftarrow{U}_i, \overrightarrow{U}_i),
\end{align}
where:
\begin{align}
    \overleftarrow{U}^t_i = e^{-i \omega^U_i t} \overleftarrow{\psi}_{\omega^U_i} \quad \text{and} \quad \overrightarrow{U}^t_i = e^{-i \omega^U_i t} \overrightarrow{\psi}_{\omega^U_i},
\end{align}
and $\psi_{\omega^U_i}^L$ and $\psi_{\omega^U_i}^R$ are defined in \eqref{eq:in_modes_left} and \eqref{eq:in_modes_right}. The set $V_i$ consists of left- and right-moving negative-energy modes:
\begin{align}
\label{not4}
    V^t_i = e^{i \omega_i^V t} G^V_i = ( \overleftarrow{V}_i, \overrightarrow{V}_i),
\end{align}
where:
\begin{align}
    \overleftarrow{V}^t_i = e^{i \omega^V_i t} \overleftarrow{\psi}^*_{-\omega^V_i} \quad \text{and} \quad \overrightarrow{V}^t_i = e^{i \omega^V_i t} \overrightarrow{\psi}^*_{-\omega^V_i}.
\end{align}
Note that we do not separate these modes into electron and positron wave functions; we only emphasize that they are orthonormal:
\begin{align}
    \langle U_i, U_j \rangle = \delta_{ij}, \quad 
    \langle V_i, V_j \rangle = \delta_{ij}, \quad \text{and} \quad 
    \langle V_i, U_j \rangle = 0,
\end{align}
and form a complete basis:
\begin{align}
\sum_i |U_i\rangle\langle U_i| + \sum_i |V_i\rangle\langle V_i| = \hat{1}.
\end{align}
Now, let us expand the solution of the Dirac equation \eqref{solution} in terms of these eigenfunctions to construct the $in$-modes:
\begin{gather}
    \psi^{u,in}_{t<-T} = u^t_i, \\ \nonumber
    \psi^{u,in}_{-T<t<T} = \sum_j \alpha^u_{ij} U^{t}_j + \beta^u_{ij} V^{t}_j, \\ \nonumber
    \psi^{u,in}_{t>T} = \sum_j A^u_{ij} u^{t}_j + B^u_{ij} v^{t}_j.
\end{gather}
At the times $t = -T$ and $t = T$, the solutions must be matched:
\begin{align}
u^{-T}_i = \sum_j \alpha^u_{ij} U^{-T}_j + \beta^u_{ij} V^{-T}_j,
\end{align}
and
\begin{align}
    \sum_j \alpha^u_{ij} U^{T}_j + \beta^u_{ij} V^{T}_j = \sum_j A^u_{ij} u^{T}_j + B^u_{ij} v^{T}_j.
\end{align}
Using the orthonormality relations for the modes, we obtain the coefficients of the $in$-mode expansion in terms of plane waves in the future:
\begin{gather}
    A^u_{ij} = \sum_k \langle u^{-T}_i, U^{-T}_k \rangle \langle U^{T}_k, u^T_j \rangle + \langle u^{-T}_i, V^{-T}_k \rangle \langle V^{T}_k, u^T_j \rangle, \\
    B^u_{ij} = \sum_k \langle u^{-T}_i, U^{-T}_k \rangle \langle U^{T}_k, v^T_j \rangle + \langle u^{-T}_i, V^{-T}_k \rangle \langle V^{T}_k, v^T_j \rangle.
\end{gather}
Similarly, for the $v$ $in$-modes:
\begin{gather}
    \psi^{v,in}_{t<-T} = v^t_i, \\
    \psi^{v,in}_{-T<t<T} = \sum_j \alpha^v_{ij} U^{t}_j + \beta^v_{ij} V^{t}_j, \\
    \psi^{v,in}_{t>T} = \sum_j A^v_{ij} u^{t}_j + B^v_{ij} v^{t}_j,
\end{gather}
we obtain:
\begin{gather}
    A^v_{ij} = \sum_k \langle v^{-T}_i, U^{-T}_k \rangle \langle U^{T}_k, u^T_j \rangle + \langle v^{-T}_i, V^{-T}_k \rangle \langle V^{T}_k, u^T_j \rangle, \\
    B^v_{ij} = \sum_k \langle v^{-T}_i, U^{-T}_k \rangle \langle U^{T}_k, v^T_j \rangle + \langle v^{-T}_i, V^{-T}_k \rangle \langle V^{T}_k, v^T_j \rangle.
\end{gather}
The $out$-modes for $t > T$ are defined as follows:
\begin{align}
   \psi^{u,out}_{t>T} = u^t_i \quad \text{and} \quad \psi^{v,out}_{t>T} = v^t_i.
\end{align}
We can now define the field operator and expand it in terms of the $in$- and $out$-modes:
\begin{align}
    \hat{\psi} = \sum_i \left( \psi^{u,in}_i a^{in}_i + \psi^{v,in}_i b_i^{in\dagger} \right) \quad \text{and} \quad \hat{\psi} = \sum_i \left( \psi^{u,out}_i a^{out}_i + \psi^{v,out}_i b_i^{out\dagger} \right).
\end{align}
Using $a^{in}_i$ and $b^{in}_i$, we define the $In$ Fock space ground state:
\begin{align}
    a^{in}_i |in \rangle = 0 \quad \text{and} \quad b^{in}_i |in\rangle = 0,
\end{align}
and using $a^{out}_i$ and $b^{out}_i$, we define the $Out$ Fock space ground state:
\begin{align}
    a^{out}_i |out \rangle = 0 \quad \text{and} \quad b^{out}_i |out\rangle = 0.
\end{align}
We then find the Bogoliubov transformation:
\begin{align}
    a^{out}_i = \sum_j \left( A^u_{ji} a^{in}_j + A^v_{ji} {b_j^{in}}^\dagger \right),
\end{align}
and
\begin{align}
    {b^{out}_i}^\dagger = \sum_j \left( B^u_{ji} a^{in}_j + B^v_{ji} {b_j^{in}}^\dagger \right).
\end{align}
Hence, the number of created particles is given by:
\begin{gather}
\label{numbpartcre}
   N = \sum_i \langle in| {a^{out}_i}^\dagger a^{out}_i |in\rangle 
   = \sum_i \sum_j {A^v_{ji}}^* A^v_{ji}. 
\end{gather}
Similarly, one can compute the number of created antiparticles, which should be equal to \eqref{numbpartcre} due to charge conservation.

We now calculate the particle current generated due to the presence of the Klein zone. First, note that when the field is turned on, a certain number of particles $N_{\text{on}}$ is created, associated with the process of switching on the external field. Similarly, when the field is turned off, $N_{\text{off}}$ particles are created. Second, after the field is turned on, a Klein zone appears, from which a current begins to flow, leading to the creation of particles and antiparticles. Since a constant current flows from the Klein zone, we expect the number of created particles to increase linearly with time: $N_{\text{Klein}} = J t$. Thus, to find the current from the Klein zone, we must calculate the linearly increasing contribution to the number of produced particles:
\begin{align}
J_{part} = \lim_{T\to \infty} \frac{N_{\text{on}} + N_{\text{off}} + N_{\text{Klein}}}{2T}.
\end{align}
Clearly, in the limit $T \to \infty$, $N_{\text{on}}$ and $N_{\text{off}}$ do not contribute to the current. A similar contribution to the total current $J_{tot}$ comes from antiparticles, $J_{anpart} = J_{part}$, so that $J_{tot} = J_{part} + J_{anpart} = 2 J_{part}$.

Separating the time-dependent part in the mode, we can rewrite the Bogoliubov coefficients using the notation from \eqref{not1}, \eqref{not2}, \eqref{not3}, and \eqref{not4}:
\begin{align}
    A^v_{ij} = e^{i (\omega^v_i - \omega^u_j) T} \sum_k \left[ e^{2 i \omega^U_k T} 
 \langle g^v_i, F^U_k \rangle \langle F^U_k, f^u_j \rangle + e^{-2 i \omega^V_k T} 
 \langle g^v_i, G^V_k \rangle \langle G^V_k, f^u_j \rangle \right].
\end{align}
Then, the number of created particles per unit time is given by:
\begin{gather}
    J_{part}=\frac{N}{2T} = \frac{1}{2T} \sum_i \sum_j {A^v_{ji}}^* A^v_{ji}
= \label{number finit time}
     \frac{1}{2T} \sum_i \sum_j \sum_k \sum_p \Bigg[ \\ \nonumber  
     e^{-2 i (\omega^U_k - \omega^U_p) T} \langle g^v_i, F^U_k \rangle^* \langle F^U_k, f^u_j \rangle^* \langle g^v_i, F^U_p \rangle \langle F^U_p, f^u_j \rangle
     + \\ \nonumber
     e^{-2 i (\omega^U_k + \omega^V_p) T} \langle g^v_i, F^U_k \rangle^* \langle F^U_k, f^u_j \rangle^* \langle g^v_i, G^V_p \rangle \langle G^V_p, f^u_j \rangle
     + \\ \nonumber
     e^{2 i (\omega^V_k + \omega^U_p) T} \langle g^v_i, G^V_k \rangle^* \langle G^V_k, f^u_j \rangle^* \langle g^v_i, F^U_p \rangle \langle F^U_p, f^u_j \rangle
     + \\ \nonumber
     e^{2 i (\omega^V_k - \omega^V_p) T} \langle g^v_i, G^V_k \rangle^* \langle G^V_k, f^u_j \rangle^* \langle g^v_i, G^V_p \rangle \langle G^V_p, f^u_j \rangle \Bigg].
\end{gather}
The last equation contains oscillating exponentials. In the limit $T \to \infty$, using the fact that $\frac{\cos(\Delta \omega T)}{T}$ vanishes, only the imaginary part of the exponential, $\frac{\sin(\Delta \omega T)}{T}$, does contribute. This quantity yields a finite current if the remaining part of the integrand is proportional to $\frac{1}{\Delta \omega}\,\delta(\Delta \omega)$. Namely:
\begin{align}\label{theconst}
    \int d\omega  \frac{\sin(\Delta \omega T)}{T}  \frac{1}{\Delta \omega} \delta(\Delta \omega) \sim \text{constant}.
\end{align}
The second and third terms in \eqref{number finit time} vanish in the limit $T \to \infty$ since $\delta(\omega_k^V + \omega_p^U) = 0$. Hence, we only need to consider the first and fourth terms in \eqref{number finit time}, which we denote as $J_{part}^1$ and $J_{part}^4$:
\begin{gather}
    J_{part}=J_{part}^1 +J_{part}^4
=
     \frac{1}{2T} \sum_i \sum_j \sum_k \sum_p \Bigg[ \\ \nonumber  
   -i  \sin\left(2 (\omega^U_k - \omega^U_p) T\right)  \langle g^v_i, F^U_k \rangle^* \langle F^U_k, f^u_j \rangle^* \langle g^v_i, F^U_p \rangle \langle F^U_p, f^u_j \rangle
     + \\ \nonumber
     i\sin\left(2  (\omega^V_k - \omega^V_p) T\right) \langle g^v_i, G^V_k \rangle^* \langle G^V_k, f^u_j \rangle^* \langle g^v_i, G^V_p \rangle \langle G^V_p, f^u_j \rangle \Bigg].
\end{gather}
Let us begin with the first term. Using the completeness relation $\sum_i |g^v_i\rangle\langle g^v_i| + \sum_i |f^u_i\rangle\langle f^u_i| = \hat{1}$, we obtain that it is given by:
\begin{gather}
 J_{part}^1=\\=\nonumber\lim_{T\to \infty}  i \frac{1}{T} \sum_i \sum_j \sum_k \sum_p  
     \sin \left(2 (\omega^U_k - \omega^U_p) T \right) \langle F^U_k, f^u_i \rangle \langle f^u_i, F^U_p \rangle \langle F^U_p, f^u_j \rangle \langle f^u_j, F^U_k \rangle
     - \\ \nonumber 
 \lim_{T\to \infty}  i \frac{1}{T} \sum_j \sum_k \sum_p  
     \sin \left(2 (\omega^U_k - \omega^U_p) T \right) \langle F^U_k, F^U_p \rangle \langle F^U_p, f^u_j \rangle \langle f^u_j, F^U_k \rangle,
\end{gather}
where the first term in the last expression vanishes because it is antisymmetric in $p$ and $k$, and the last term is nonzero since $\langle F^U_k, F^U_p \rangle = \delta_{kp}$. Namely, the nonzero contribution to the current is contained in the expression:
\begin{gather}
J_{part}^1= \lim_{T\to \infty}  -i \frac{1}{T} \sum_j \sum_k \sum_p  
     \sin \left(2 (\omega^U_k - \omega^U_p) T \right) \langle F^U_k, F^U_p \rangle \langle F^U_p, f^u_j \rangle \langle f^u_j, F^U_k \rangle.
\end{gather}
The same applies to the last term $J_{part}^4$ in Eq. (\ref{number finit time}). As a result, we obtain a compact expression for the particle contribution to the current:
\begin{gather}
J_{part} = \lim_{T\to \infty}  -i \frac{1}{2T} \sum_j \sum_k \sum_p  
     \Bigg[  
     \sin \left(2 (\omega^U_k - \omega^U_p) T \right) \langle F^U_k, F^U_p \rangle \langle F^U_p, f^u_j \rangle \langle f^u_j, F^U_k \rangle
     - \\ \nonumber
     - \sin \left(2 (\omega^V_k - \omega^V_p) T \right) \langle G^V_k, G^V_p \rangle \langle G^V_p, f^u_j \rangle \langle f^u_j, G^V_k \rangle \Bigg].
\end{gather}
Changing the variables $p$ and $k$, we obtain:
\begin{gather}
\label{current 2}
J_{part} = \lim_{T\to \infty}  -i \frac{1}{4T} \sum_j \sum_k \sum_p  
     \Bigg[ \\ \nonumber 
     \sin \left(2 (\omega^U_k - \omega^U_p) T \right) \langle F^U_k, F^U_p \rangle \left( \langle F^U_p, f^u_j \rangle \langle f^u_j, F^U_k \rangle - \langle F^U_k, f^u_j \rangle \langle f^u_j, F^U_p \rangle \right)
     - \\ \nonumber
     \sin \left(2 (\omega^V_k - \omega^V_p) T \right) \langle G^V_k, G^V_p \rangle \left( \langle G^V_p, f^u_j \rangle \langle f^u_j, G^V_k \rangle - \langle G^V_k, f^u_j \rangle \langle f^u_j, G^V_p \rangle \right) \Bigg].
\end{gather}
Using $\langle F^U_k, F^U_p \rangle = \delta_{kp} = \delta(\omega^U_k - \omega^U_p)$ and $\langle G^V_k, G^V_p \rangle = \delta_{kp} = \delta(\omega^V_k - \omega^V_p)$, we obtain expressions containing an integral of the form (\ref{theconst}), which gives a finite contribution to the current in the limit $T \to \infty$. A straightforward but lengthy calculation shows that only modes from the Klein zone contribute to the current, which takes the following form:
\begin{align}
   \boxed{ J_{tot} = \int_{m}^{V - m} \frac{d\omega}{2\pi} \frac{4\kappa_{\omega}}{(1 + \kappa_{\omega})^2} }.
\end{align}
This expression coincides with the current obtained in the previous sections up to a sign, which cannot be reproduced within the approach used in the present section.

\section{Conclusions and acknowledgments}

Thus, in the presence of a constant electric potential with different asymptotic values at spatial infinity, the definition of a particle becomes ambiguous. Depending on the choice of mode basis used for quantization, the corresponding Fock space ground states yield different expectation values for the current operator when the potential is strong enough to allow pair creation. For one specific choice of modes, the current is zero.

To resolve this ambiguity, we consider scenarios in which the background potential is rapidly switched on at a specific time. In this case, the notion of a particle is unambiguous, being defined by the mode basis that diagonalizes the free Hamiltonian in empty Minkowski spacetime before the field is turned on. Under these conditions, the current at future infinity is both nonzero and time independent.

We emphasize that the expectation value of the current at future infinity is the same, regardless of whether the potential remains constant after being switched on or is eventually switched off after a long time. As explained following Eq. (\ref{eq:ground_state_definition}), this current is time independent because the system is considered in an infinite volume.

However, the situation changes if the system is confined to a finite-sized box. The electron–positron pairs created by the electric field will gradually fill the discrete energy levels until the electric current vanishes. To see such a process one has to consider 4D interacting theory rather than 2D and gaussian one. But as the size of the box increases to infinity, the time required for the system to reach this equilibrium state also diverges, thereby recovering the time-independent current of the infinite-volume case.

We would like to acknowledge discussions with S.Semenov. The work of D.Diakonov and D.Sadekov was supported by the Foundation for the Advancement of Theoretical Physics and Mathematics ``BASIS''. This work was partially supported by the Ministry of Science and Higher Education of the Russian Federation (agreement no. 075–15–2022–287).

\newpage
\begin{appendices} 
\setcounter{equation}{0}
\renewcommand\theequation{A.\arabic{equation}}

\section{Free Dirac equation}
The free Dirac equation in $2D$ is given by:
\beq
\label{deq}
    \Big(i\gamma^{\mu}\partial_{\mu}-m\Big)\psi(t,x)=0,
\eeq
\beq
    \gamma^{0} = \begin{pmatrix}
	1 & 0\\
	0 & -1
	\end{pmatrix}, \; \gamma^{1} = \begin{pmatrix}
	0 & i\\
	i & 0
	\end{pmatrix}.
\eeq 
Solutions of (\ref{deq}) with fixed frequency are given in the form $\psi_{\omega}(t,x)=e^{-i\omega t}\psiw(x) = e^{-i\omega t}\begin{pmatrix}
    \psi_L(x)\\
    \psi_R(x)
    \end{pmatrix}$. In terms of the components $\psi_L,\psi_R$ eq.\eqref{deq} reduces to: 
\begin{equation}
	\begin{cases}
	\big(\omega -m\big)\psi_L(x) - \partial_x \psi_R(x) = 0,\\
	\big(\omega +m\big)\psi_R(x) + \partial_x \psi_L(x) = 0.
	\end{cases}
\end{equation}    
For each frequency $\omega\in(-\infty,m)\cup (m,\infty)$ there are two solutions --- the right-moving waves:
\begin{gather}
\label{left free}
    \psiwL(x) = \frac{1}{\sqrt{2\kw\left|\omega-m\right|}}
    \begin{pmatrix}
        -i\kw \\
        \omega -m
    \end{pmatrix}e^{-i\kw x} 
\end{gather}
and the left-moving ones:
\begin{gather}
\label{right free}
    \psiwR(x) = \frac{1}{\sqrt{2\kw\left|\omega-m\right|}}
    \begin{pmatrix}
        i\kw \\
        \omega -m
    \end{pmatrix}e^{i\kw x} ,
\end{gather}
where $\kw=\sqrt{\omega^2-m^2}$ .

Having at hand the orthonormal set of modes, we can expand the fermion field operator over the modes in the form:

\begin{gather}
\label{operator free dirac}
\widehat{\psi}(t,x) 
=\\=
\nonumber
 \int_{m}^\infty \frac{d \omega}{2\pi}  \left(\psiwL(x) e^{-i\omega t} \hataL_{\w} + \psiwR(x) e^{-i\omega t} \hataR_{\w}\right)
+\\+ \nonumber
\int_m^\infty \frac{d \omega}{2\pi}  \left(\overleftarrow{\psi}^*_{-\omega}(x) e^{i\omega t} \hatbL\phantom{\empty}^{\dagger}_{\w} +\overrightarrow{\psi}^*_{-\omega}(x) e^{i\omega t} \hatbR\phantom{\empty}^{\dagger}_{\w} \right).
\end{gather}
In the main body of the paper we use these modes to construct $In$ and $Out$ states, when the background field is switched on and off.

\section{Single-particle solution}\label{app:1_particle_solution}
The solutions (\ref{eq:in_modes_left})–(\ref{eq:in_modes_right}) can be regarded as electron wavefunctions with a continuous probability current. Consider, for example, the right-moving solutions (\ref{eq:in_modes_right}). In the Klein zone with $\chi(\w)=-1$ the currents for the incident, transmitted, and reflected parts are, correspondingly, given by
\beq
    j_{in} = \left|B_{\w}\right|^2 
    \begin{pmatrix}
        i\kw\\
        \omega -m
    \end{pmatrix}^{\dagger} \gamma^0\gamma^1\begin{pmatrix}
        i\kw\\
        \omega -m
    \end{pmatrix} = \sign\left(\w-m\right)=1\;,
\eeq
\beq
    j_{t} = \left|B_{\w}\right|^2 \cdot\frac{4\kw^2}{\pw^2\left(\kappaw+1\right)^2}\cdot
    \begin{pmatrix}
        -i\pw\\
        \omega -V_0-m
    \end{pmatrix}^{\dagger} \gamma^0\gamma^1\begin{pmatrix}
        -i\pw\\
        \omega -V_0-m
    \end{pmatrix} = \frac{4\kappaw}{\left(\kappaw+1\right)^2}\;,
\eeq
\beq
    j_{r} = -\left|B_{\w}\right|^2 \cdot\frac{\left(\kappaw-1\right)^2}{\left(\kappaw+1\right)^2}\cdot
    \begin{pmatrix}
        -i\kw\\
        \omega -m
    \end{pmatrix}^{\dagger} \gamma^0\gamma^1\begin{pmatrix}
        -i\kw\\
        \omega -m
    \end{pmatrix} = -\frac{\left(\kappaw-1\right)^2}{\left(\kappaw+1\right)^2}\;.
\eeq
With our definitions (\ref{eq:notations}) the kinematic factor $\kappaw$ is, as it should be, positive, so the directions of the probability currents correspond to the physical interpretation for an incoming particle, despite the fact that the current and momentum of the transmitted part are opposite in the Klein zone.

\end{appendices}

\bibliography{literature}
\bibliographystyle{unsrt}
\end{document}

%% file: Defs.tex
\def\kw{k_{\omega}} \def\pw{p_{\omega}}
\def\kappaw{\varkappa_{\omega}}
\def\psiw{\psi_{\omega}}
\def\psiwR{\overrightarrow{\psi}_{\omega}}
\def\psiwL{\overleftarrow{\psi}_{\omega}}
\def\hataL{\widehat{ \overset{\smash{\raisebox{-0.45ex}{$\scriptstyle\leftarrow$}}}{a} }}
\def\hatbL{\widehat{ \overset{\smash{\raisebox{-0.5ex}{$\scriptstyle\leftarrow$}}}{b} }}
\def\hataR{\widehat{ \overset{\smash{\raisebox{-0.45ex}{$\scriptstyle\rightarrow$}}}{a} }}
\def\hatbR{\widehat{ \overset{\smash{\raisebox{-0.5ex}{$\scriptstyle\rightarrow$}}}{b} }}
\def\w{\omega}
\def\I{\mathrm{I}}
\def\II{\mathrm{II}}
\def\III{\mathrm{III}}
\def\IV{\mathrm{IV}}
\def\V{\mathrm{V}}
\def\sign{\text{sign}}
\def\kwh{k_{\omega'}} \def\pwh{p_{\omega'}}
\def\kappawh{\varkappa_{\omega'}}
\def\hatcL{\widehat{ \overset{\smash{\raisebox{-0.45ex}{$\scriptstyle\leftarrow$}}}{c} }}
\def\hatdL{\widehat{ \overset{\smash{\raisebox{-0.5ex}{$\scriptstyle\leftarrow$}}}{d} }}
\def\hatcR{\widehat{ \overset{\smash{\raisebox{-0.45ex}{$\scriptstyle\rightarrow$}}}{c} }}
\def\hatdR{\widehat{ \overset{\smash{\raisebox{-0.5ex}{$\scriptstyle\rightarrow$}}}{d} }}